\pdfoutput=1
\RequirePackage{ifpdf}
\ifpdf 
\documentclass[pdftex]{sigma}
\else
\documentclass{sigma}
\fi

\numberwithin{equation}{section}

\begin{document}

\allowdisplaybreaks

\newcommand{\arXivNumber}{1603.03528}

\renewcommand{\PaperNumber}{112}

\FirstPageHeading

\ShortArticleName{Integrability of Nonholonomic Heisenberg Type Systems}

\ArticleName{Integrability of Nonholonomic\\ Heisenberg Type Systems}

\Author{Yury A.~GRIGORYEV~$^\dag$, Alexey P.~SOZONOV~$^\dag$ and Andrey V.~TSIGANOV~$^{\dag\ddag}$}

\AuthorNameForHeading{Yu.A.~Grigoryev, A.P.~Sozonov and A.V.~Tsiganov}

\Address{$^\dag$~St.~Petersburg State University, St.~Petersburg, Russia}
\EmailD{\href{mailto:yury.grigoryev@gmail.com}{yury.grigoryev@gmail.com}, \href{mailto:sozonov.alexey@yandex.ru}{sozonov.alexey@yandex.ru}, \href{mailto:andrey.tsiganov@gmail.com}{andrey.tsiganov@gmail.com}}

\Address{$^\ddag$~Udmurt State University, Izhevsk, Russia}

\ArticleDates{Received March 17, 2016, in f\/inal form November 22, 2016; Published online November 25, 2016}

\Abstract{We show that some modern geometric methods of Hamiltonian dynamics can be directly applied to the nonholonomic Heisenberg type systems. As an example we present characteristic Killing tensors, compatible Poisson brackets, Lax matrices and classical $r$-matrices for the conformally Hamiltonian vector f\/ields obtained in a process of reduction of Hamiltonian vector f\/ields by a nonholonomic constraint associated with the Heisenberg system.}


\Classification{37J60; 70G45; 70H45}

\section{Introduction}
Hamiltonian mechanics emerged in 1833 as a convenient reformulation of classical Newtonian mechanics and analytical Lagrangian mechanics. Over the next almost two centuries, it was gradually realized that Hamiltonian formulation of a system of dif\/ferential equations
\begin{gather*}
\frac{d}{dt}x_k=X_k,\qquad X=PdH
\end{gather*}
has several advantages based on the main postulate that evolution of a physical system over time is governed by a~single Hamiltonian $H$ of that system and a Poisson bivector $P$ describing geometry and topology of the phase space. Many dif\/ferent mathematical methods and concepts are widely used in the Hamiltonian formalism. For instance, we can construct f\/irst integrals of the Hamiltonian vector f\/ield $X$ using Killing tensors (Riemannian geometry), compatible Poisson brackets (bi-Hamiltonian geometry) or Lax matrices (algebraic geometry, Lie algebras theory, classical $r$-matrix theory etc.).

It is natural to ask what happens to these mathematical methods when we impose nonholonomic constraints on the Hamiltonian system and make a suitable reduction.
Usually presence of the constraints drastically modif\/ies or destroys these geometric constructions even when one gets a conformally Hamiltonian vector f\/ield
\begin{gather*}
\hat {X}=\mu(x)\hat{P}d\hat{H}
\end{gather*}
after the reduction, see \cite{bates93,bbm15,bl03,bm15,bmb13,bm14,cus10,fas15,ts12,mash94} and references within. Here~$\hat{H}$ and~$\hat{P}$ are Hamiltonian and Poisson bivector on reduced phase space, whereas function $\mu(x)$ is the so-called conformal factor.

The main aim of this note is to present a family of conformally Hamiltonian dynamical systems for which all the geometric methods listed above can be applied without any additional modif\/ications. The corresponding nonholonomic constraint is associated with the so-called Heisenberg system or nonholonomic integrator, which plays an important role in both nonlinear control and nonholonomic dynamics \cite{bl03}. We have to underline that the integrable systems discussed in this note look quite artif\/icial from the viewpoint of control theory and mechanics but they provide an example of the standard geometric methods applicability to the conformally Hamiltonian systems.

This paper is organized as follows. Section~\ref{section1.1} recalls a brief description of the constrained motion in three-dimensional Euclidean space. Section~\ref{section2} contains the main results on reduced motion of the nonholonomic Heisenberg type systems on the plane. We will show that integrable potentials for this non Hamiltonian vector f\/ield satisfy to the Bertrand--Darboux type equation. Solutions of this equation and the corresponding characteristic coordinates will be explicitly determined. Section~\ref{section3} deals with application of the standard St\"{a}ckel theory to the conformally Hamiltonian vector f\/ield. We will discuss a construction of the St\"{a}ckel matrices, compatible Poisson structures, Lax matrices and classical $r$-matrices.

\subsection{Main def\/initions}\label{section1.1}
According to \cite{bl03,mol12,mash94} we take the standard Hamiltonian equations of motion in Euclidean space $\mathbb R^3$
\begin{gather}\label{eq-3dh}
 \dot{q_i}=\frac{\partial H}{\partial p_i},\qquad \dot{p}_i=-\frac{\partial H}{\partial q_i},\qquad i=1,2,3,
\end{gather}
where
\begin{gather}\label{h-free}
H=\frac{1}{2}\big(p_1^2+p_2^2+p_3^2\big)+V(q_1,q_2,q_3).
\end{gather}
On the phase space $M=T^*\mathbb R^3$ we can introduce coordinates $x=(q,p)$ in which equations (\ref{eq-3dh}) have form $\dot{x}_i=X_i$ and to determine Hamiltonian vector f\/ield
\begin{gather*}
X=\sum_{i=1}^{6} X_i\frac{\partial}{\partial x_i},
\end{gather*}
which is a linear operator on a space of the smooth functions on $M$ that encodes the evolution of any quantity
\begin{gather*}
\dot{F}=X(F)=\sum_{i=1}^{6} X_i \frac{\partial F}{\partial x_i}.
\end{gather*}
Let us impose the constraint of f\/irst order in momenta (velocities)
\begin{gather}\label{f-gen}
 f=(b,p)=0,
\end{gather}
where $b=(b_1,b_2,b_3)$ is a vector depending on coordinates $q$ and $(x,y)$ means an inner product in $\mathbb R^3$. In this case equations of motion are written in the following form
\begin{gather}\label{eq-3d}
 \dot{q_i}=\frac{\partial H}{\partial p_i},\qquad \dot{p}_i=-\frac{\partial H}{\partial q_i}+\lambda b_i,\qquad i=1,2,3,
\end{gather}
together with the constraint equation (\ref{f-gen}). The corresponding vector f\/ield looks like an additive perturbation of the initial Hamiltonian vector f\/ield
\begin{gather*}\hat{X}= X\big|_{b=0}+\lambda \left(b_1\frac{\partial }{\partial p_1}+b_2\frac{\partial }{\partial p_2}+b_3\frac{\partial }{\partial p_3}\right),\end{gather*}
where unknown Lagrange multiplier $\lambda$ has to be computed from the condition
\begin{gather*}\dot{f}=\hat{X}(f)=X(f)\big|_{b=0}+\lambda (b,b)=0,\end{gather*}
such that
 \begin{gather}\label{l-mul}
 \lambda=\frac{X(f)|_{b=0}}{(b,b)}.
 \end{gather}
Here $X(f)|_{b=0}$ denotes the vector f\/ield in the absence of constraint. Such equations and an equivalence of Hamiltonian and Lagrangian reductions are carefully discussed in the book~\cite{bl03}.

For completeness and self-suf\/f\/iciency of presentation we consider integrable constraint
\begin{gather}\label{f-hol}
f=p_3=0,\qquad b=(0,0,1),
\end{gather}
when a third component of momenta is equal to zero, and non integrable constraint
\begin{gather}\label{f-nhol}
f=p_3-(q_2p_1-q_1p_2)=0,\qquad b=(-q_2,q_1,1),
\end{gather}
when third components of momenta and angular momenta coincide with each other.

If we assume that the potential $V$ in $\mathbb R^3$ does not depend on $q_3$, then Lagrange multipliers~(\ref{l-mul}) are equal to
\begin{gather*}
\lambda=0\qquad\mbox{and}\qquad\lambda=\frac{q_1\partial_2V-q_2\partial_1 V}{1+q_1^2+q_2^2},\qquad \partial_j=\frac{\partial}{\partial q_j},
\end{gather*}
respectively. Thus, the Hamiltonian system (\ref{h-free}) with integrable constraint~(\ref{f-hol}) represent a Hamiltonian system on the plane $\mathbb R^2$ embedded in a three dimensional Euclidean space~$\mathbb R^3$ in which there is no force acting on the third component. If we impose non integrable constraint~(\ref{f-nhol}), the equations for~$q_3$ also decouples from the rest of the system~(\ref{eq-3d}) and we obtain a~two-degrees of freedom non-Hamiltonian system on the plane. Following to~\cite{bl03,mol12} we will call dynamical system associated with $\hat{X}$ as a~Heisenberg type system.

Integrability of the Hamiltonian systems on the plane is the well-known problem of classical mechanics considered by Bertrand~\cite{bert} and Darboux~\cite{darb}. Later on results of this investigation due to Darboux were included in the textbook of Whittaker on analytical mechanics~\cite{witt37} almost verbatim. In the next Section we compare classical Bertrand--Darboux theorem with its non-Hamiltonian counterpart, which appears in the nonholonomic case.

\section{Reduced systems on the plane}\label{section2}
When $V=V(q_1,q_2)$, substituting integrable constraint $p_3=0$ (\ref{f-hol}) into (\ref{h-free}) one gets
\begin{gather}\label{hh}
H_1=\frac{1}{2}\bigl(p_1^2+p_2^2\bigr)+V(q_1,q_2)=\frac{1}{2} \mathrm g^{ij}p_ip_j+V(q_1,q_2),\qquad \mathrm g=\left(
 \begin{matrix}
 1 &0 \\
 0 & 1
 \end{matrix}
 \right)
\end{gather}
with the summation convention in force; we regard the $\mathrm g^{ij}$ as components of the covariant form of the metric tensor $\mathrm g$ on the plane, and use this metric freely to raise and lower indices.

After standard Hamiltonian reduction by cyclic third coordinate, the original Hamiltonian vector f\/ield (\ref{eq-3dh}) in $T^*\mathbb R^3$ becomes the Hamiltonian vector f\/ield in $T^*\mathbb R^2$
\begin{gather}\label{conf-Xh}
\dot{x}=X,\qquad X=P dH_1,\qquad P=\left(
 \begin{matrix}
 0 & I \\
 -I & 0
 \end{matrix}
 \right),
\end{gather}
where $x=(q_1,q_2,p_1,p_2)$. The Poisson bivector $P$ in (\ref{conf-Xh}) def\/ines canonical Poisson bracket
\begin{gather}\label{can-br}
\{q_i,p_i\}=1,\qquad \{q_i,p_k\}= \{p_1,p_2\}=0,\qquad i\neq k.
\end{gather}
Of course, this vector f\/ield $X$ preserves Poisson structure $P$, energy $H_1$ and the standard volume 2-form $\Omega=dq\wedge dp$.

Substituting non integrable constraint $p_3=q_2p_1-q_1p_2$ (\ref{f-nhol}) into (\ref{h-free}) one gets
\begin{gather}
\hat{H}_1=\frac{1}{2}\bigl(p_1^2+p_2^2+(q_2p_1-q_1p_2)^2\bigr)+V(q_1,q_2) =\frac{1}{2} \hat{\mathrm g}^{ij}p_ip_j+V(q_1,q_2),\label{h-non}\\
\hat{\mathrm g}=\left( \begin{matrix}
 1+q_2^2 & -q_1q_2 \\
 -q_1q_2 & 1+q_1^2
 \end{matrix}
 \right).\nonumber
\end{gather}
The second order contravariant symmetric tensor f\/ield $\hat{\mathrm g} $ is a Killing tensor with respect to the standard metric tensor $\mathrm g$, i.e., it satisf\/ies to the Killing equation
\begin{gather*}
[\![\mathrm g,\hat{\mathrm g}]\!]=0,
\end{gather*}
where $[\![\cdot,\cdot ]\!]$ is a Schouten bracket. It allows us to construct conformal Killing tensor of second order
\begin{gather*}
G=\hat{\mathrm g}-\operatorname{tr}\left(\hat{\mathrm g}\right) \mathrm g=-\left(
 \begin{matrix}
 1+q_1^2 & q_1q_2 \\
 q_1q_2 & 1+q_2^2
 \end{matrix}
 \right)
\end{gather*}
 with vanishing Nijenhuis torsion and to def\/ine the standard Turiel deformation of the standard symplectic form
\begin{gather*}
\hat{\omega}=d\big( G^{ij} p_idq_j\big),
\end{gather*}
see \cite{ts11,ts12} and references within. The corresponding Poisson bracket
\begin{gather}\label{nhcan-br}
\{q_i,p_i\}^\wedge=1+q_i^2,\qquad \{q_i,p_j\}^\wedge=q_1q_2,\qquad \{p_1,p_2\}^\wedge=q_1p_2-q_2p_1,\qquad i\neq j
\end{gather}
is compatible with canonical bracket (\ref{can-br}). Because the Turiel deformation is trivial deformation in the Lichnerowicz--Poisson sense, there is a~change of variables
\begin{gather}\label{n-var}
p_1\to\pi_1=\frac{q_1q_2p_2-\big(1+q_2^2\big)p_1}{1+x_1^2+x_2^2},\qquad
p_2\to\pi_2=\frac{q_1q_2p_1-\big(1+q_1^2\big)p_2}{1+x_1^2+x_2^2},
\end{gather}
which transforms this bracket to the canonical one (\ref{can-br}), similar to other nonholonomic systems with constraints of f\/irst order in momenta~\cite{bts12,bm14,ts11a}.

After the nonholonomic reduction, original Hamiltonian vector f\/ield (\ref{eq-3dh}) in $T^*\mathbb R^3$ becomes a conformally Hamiltonian vector f\/ield in $T^*\mathbb R^2$
\begin{gather}\label{conf-X}
\dot{x}=\hat{X},\qquad \hat{X}=\mu \hat{P} d\hat{H}_1,\qquad \hat{P}= \hat{\omega}^{-1},
\end{gather}
where conformal factor
\begin{gather*}
\mu=\big(1+q_1^2+q_2^2\big)^{-1}
\end{gather*}
is a nowhere vanishing smooth function on the plane. This vector f\/ield possesses energy~$\hat{H}_1$ and the volume 2-form $\hat{\Omega}=\mu dq\wedge dp$, but it does not preserve the Poisson bivector~$\hat{P}$.

\subsection{Bertrand--Darboux type equation}
 An existence of the integrals of second order in velocities for Hamiltonian vector f\/ield (\ref{conf-Xh}) is described by a classical Bertrand--Darboux theorem~\cite{bert,darb,witt37}.
 \begin{proposition}
The function \eqref{hh}
 \begin{gather}\label{h1-hol}
 H_1=\frac{1}{2} \mathrm g^{ij} p_ip_j+V(q_1,q_2),\qquad \mathrm g=\left(
 \begin{matrix}
 1 & 0 \\
 0 & 1
 \end{matrix}
 \right)
 \end{gather}
defines integrable Hamiltonian vector field $X$ \eqref{conf-Xh}, which has an independent quadratic first integral
 \begin{gather}\label{h2-hol}
 H_2= K^{ij}p_ip_j+U(q_1,q_2),
 \end{gather}
if the second order contravariant symmetric tensor field~$K$ obeys the Killing tensor equation
\begin{gather}\label{meq-1h}
[\![\mathrm g,K]\!]=0,
\end{gather}
 and potential $V$ satisfies the compatibility condition
\begin{gather}\label{meq-2h}
 d(\mathbf KdV)=0,\qquad \mathbf{K}=K \mathrm g^{-1},
\end{gather}
where $\mathbf K$ is the tensor field of $(1,1)$ type.

A characteristic coordinate system for \eqref{meq-2h} provides separation for the potential~$V$ and can be taken as one of the following four orthogonal coordinate systems on the plane: elliptic, parabolic, polar or Cartesian.
 \end{proposition}

If a non-trivial solution exists, the Bertrand--Darboux equation (\ref{meq-2h}) can be reduced to canonical form by transforming to characteristic coordinates, which appear to be separation coordinates for the Hamilton--Jacobi equation related to the natural Hamiltonian (see Bertrand~\cite{bert}, Darboux~\cite{darb}, or Whittaker~\cite{witt37} and Ankiewicz and Pask~\cite{ap83} for a full proof of the BD theorem). The modern proof of this classic statement may be found in \cite{sm08}.

In nonholonomic case we can also substitute the same ansatz (\ref{h2-hol}) into the equation $\dot{H}_2=0$ and get the following generalisation of the Bertrand--Darboux result.
 \begin{proposition}
The function \eqref{h-non}
 \begin{gather}\label{h1-nhol}
\hat{H}_1=\frac{1}{2} \hat{\mathrm g}^{ij} p_ip_j+V(q_1,q_2),\qquad\hat{\mathrm g}=\left(
 \begin{matrix}
 1+q_2^2 & -q_1q_2 \\
 -q_1q_2 & 1+q_1^2
 \end{matrix}
 \right)
 \end{gather}
defines integrable conformally Hamiltonian vector field $\hat{X}$ \eqref{conf-X}, which has an independent quad\-ra\-tic first integral
 \begin{gather}\label{h2-nhol}
 \hat{H}_2= K^{ij}p_ip_j+U(q_1,q_2),
 \end{gather}
if the second order contravariant symmetric tensor field $K$ is a Killing tensor
\begin{gather}\label{meq-1}
[\![\mathrm g,K]\!]=0,\qquad \mathrm g=\left(
 \begin{matrix}
 1 & 0 \\
 0 & 1
 \end{matrix}
 \right)
\end{gather}
and potential $V$ satisfies the compatibility condition
\begin{gather}\label{meq-2}
 d\big( \hat{\mathbf K}dV\big)=0,\qquad \hat{\mathbf K}=K \hat{\mathrm g}^{-1}.
\end{gather}
A characteristic coordinate system for \eqref{meq-2} provides separation for potential $V(q_1,q_2)$.
 \end{proposition}

In this case existence of the second integral of motion guarantees integrability of the given non Hamiltonian vector f\/ield $\hat{X}$ according to the Euler--Jacobi theorem~\cite{koz13}. Equation~(\ref{meq-2}) was brief\/ly discussed in~\cite{ts15b} using coordinates $q_{1,2}$ and momenta~$\pi_{1,2}$~(\ref{n-var}).

\begin{remark}\sloppy
In contrast with the standard Bertrand--Darboux theorem the Killing equa\-tion~(\ref{meq-1}) and compatibility condition~(\ref{meq-2}) include two dif\/ferent metrics $\mathrm g$ and $\hat{\mathrm g}$. It happens because the reaction force is entered in the second part of the vector f\/ield (\ref{eq-3d}) only. Sequentially, $K$~is a Killing tensor with respect to the Euclidean metric $\mathrm g$ to the plane, which is an induced metric associated with standard embedding $\mathbb R^2\subset \mathbb R^3$ in Cartesian coordinates. However, $K$ is not a Killing tensor with respect to metric $\hat{\mathrm g}$, which appears in the reduced Hamiltonian~(\ref{h-non}) after nonholonomic reduction and then def\/ines raising and lowering indices in the compatibility condition~(\ref{meq-2}).
\end{remark}

\subsection{Solutions of the Bertrand--Darboux type equation}
In~\cite{darb} Darboux proceeds to f\/ind the unknown potential $V(q_1,q_2)$ by solving the compatibility condition~(\ref{meq-2h}) using method of characteristics. Solving the second compatibility condition~(\ref{meq-2}) we can apply the same method.

In both cases characteristic coordinates for (\ref{meq-2h}) and (\ref{meq-2}) consist of eigenvalues of the corresponding $(1,1)$ tensor f\/ield ${\mathbf K}$ or $\hat{\mathbf K}$. In order to describe these coordinates we can start with a well-known generic solution of the common Killing tensor equations~(\ref{meq-1h}) and~(\ref{meq-1})
\begin{gather}\label{gen-K}
K=\left( \begin{matrix}
 c_1 q_2^2+2 c_2 q_2+ c_3 &- c_1 q_1 q_2-c_2 q_1- c_4 q_2 +c_6\\
 - c_1 q_1 q_2-c_2 q_1- c_4 q_2+c_6 & c_1 q_1^2+ 2c_4 q_1+ c_5
 \end{matrix}
 \right),
\end{gather}
which depends on six constants of integration $c_1,\ldots,c_6$.

Substituting this generic solution into the standard compatibility condition (\ref{meq-2h}) one gets
\begin{gather}\label{bd-eq}
\bigl(A\partial_{11}+B\partial_{22}+C\partial_{12} +a\partial_1+b\partial_2\bigr)V(q_1,q_2)=0,
 \end{gather}
where $\partial_i={\partial}/{\partial q_i}$, $ \partial_{ik} ={\partial^2}/{\partial q_i\partial q_k}$ and
polynomials of second order read as
\begin{gather*}
A=c_1 q_1q_2+c_2 q_1+c_4 q_2-c_6,\!\!\qquad B=-A,\!\!\qquad C=c_1 q_2^2-c_1
q_1^2+2c_2 q_2-2c_4 q_1+c_3-c_5,
\end{gather*}
whereas polynomials of the f\/irst order are
 \begin{gather*}
 a=3(c_1 q_2+c_2 ),\qquad b=-3(c_1q_1+c_4 ).
 \end{gather*}
 The linear PDE (\ref{bd-eq}) of second order was obtained by Bertrand \cite{bert} and studied by Darboux~\cite{darb}. Thus, it was later called the Bertrand--Darboux equation~\cite{sm08,witt37}.

Solving (\ref{bd-eq}) for $V (q_1,q_2)$ amounts to f\/inding admissible potentials of the Hamiltonian systems def\/ined by~$H$~(\ref{hh}), which integrability is af\/forded by the existence of f\/irst integrals~(\ref{h2-hol}) which are quadratic in the momenta. Second potential $U(q_1,q_2)$ in~(\ref{h2-hol}) is a~solution of the equation
\begin{gather*}
d(\mathbf KdU)=V.
\end{gather*}

In nonholonomic case, substituting the same generic solution (\ref{gen-K}) into the compatibility condition (\ref{meq-2}) one gets similar linear PDE
\begin{gather}\label{bd-non}
\big(1+q_1^2+q_2^2\big)\big(\hat{A}\partial_{11}+\hat{B}\partial_{22}+\hat{C}\partial_{12}\big)V(q_1,q_2)
+\big(\hat{a}\partial_1+\hat{b}\partial_2\big)V(q_1,q_2)=0,
\end{gather}
 with coef\/f\/icients
\begin{gather*}
\hat{A}= \big(1+q_1^2\big)(q_1c_2-c_6)+q_2\big(1-q_1^2\big)c_4+q_1q_2(c_1-c_5),\\
\hat{B}=-\big(1+q_2^2\big)(q_2c_4-c_6) -q_1\big(1-q_2^2\big)c_2-q_1q_2(c_1-c_3),\\
\hat{C}=-\big(q_1^2-q_2^2\big)c_1+\big(1+q_1^2\big)(2q_2c_2+c_3)-\big(1+q_2^2\big)(2q_1c_4+c_5),
\end{gather*}
and
\begin{gather*}
\hat{a} = \big(q_1^2+q_2^2+3\big) \big(c_1 q_2+2 c_2 q_1^2-2 c_4 q_1 q_2\big)-q_2\big(2\big(q_1^2+1\big) c_3-\big(q_1^2-q_2^2-1\big) c_5\big)\\
\hphantom{\hat{a} =}{} +q_1 \big(q_1^2-3 q_2^2+1\big) c_6-\big(q_1^2+q_2^2-3\big) c_2,\\
\hat{b} =-\big(q_1^2+q_2^2+3\big) \big(c_1 q_1+2 c_4 q_2^2-2 c_2 q_1 q_2\big)+q_1\big(2 \big(q_2^2+1\big)c_5+ \big(q_1^2-q_2^2+1\big) c_3\big)\\
\hphantom{\hat{b} =}{} +q_2 \big(3 q_1^2-q_2^2-1\big) c_6+\big(q_1^2+q_2^2-3\big) c_4.
\end{gather*}
 Solutions $V(q_1q_2)$ of this Bertrand--Darboux type equation (\ref{bd-non}) determine all the admissible Jacobi integrals~(\ref{h-non}), which def\/ine integrable conformally Hamiltonian vector f\/ields~(\ref{conf-X}) with f\/irst integrals~(\ref{h2-nhol}) of second order in momenta. In this case potential $U(q_1,q_2)$ in~(\ref{h2-nhol}) is a~solution of the equation
\begin{gather*}
d\big(\hat{\mathbf K}dU\big)=V.
\end{gather*}
\begin{remark}
Before solving the PDE (\ref{bd-eq}), Darboux ingeniously observes that it can be simplif\/ied without loss of generality \cite{darb}. Indeed, by rotating and translating the axes, one can simplify the general solution of the Killing tensor equation, thus bringing it to a certain canonical form. In modern language solving the equivalence and canonical forms problem for the Killing tensor equations~(\ref{meq-1h}), (\ref{meq-1}) is equivalent to analysing the orbits of the six-dimensional vector space of solutions under the action of the Lie group of orientation-preserving isometries.
\end{remark}
It is well known that there are four types of orbits generated by the following canonical Killing tensors:
\begin{alignat}{3}
& \text{Cartesian:} \quad &&
K^{(1)}=\left(
 \begin{matrix}
 1 & 0 \\
 0 & 0
 \end{matrix}
 \right) ,&\nonumber\\
& \text{polar:}\quad &&
K^{(2)}=\left(
 \begin{matrix}
 q_2^2 & -q_1q_2 \\
 -q_1q_2 & q_1^2
 \end{matrix}
 \right),& \nonumber\\
& \text{parabolic:}\quad &&
K^{(3)}=\left(
 \begin{matrix}
 0 & -q_2 \\
 -q_2 & 2q_1 \\
 \end{matrix}
 \right), & \nonumber\\
& \text{elliptic:}\quad &&
K^{(4)}=\left(
 \begin{matrix}
 \kappa^2+ q_2^2 & -q_1q_2 \\
 -q_1q_2 & q_1^2
 \end{matrix}
 \right),\qquad \kappa\in\mathbb R, & \label{4-ten}
\end{alignat}
associated with Cartesian, polar, parabolic and elliptic coordinate systems on the plane, respectively.

Let us consider these characteristic coordinate systems associated with the Killing tensors~(\ref{4-ten}).

\textit{Elliptic coordinates.}
Substituting Killing tensor $K^{(4)}$ into the (1,1) tensor $\mathbf K= K^{(4)} \mathrm g^{-1}$ and calculating its eigenvalues one gets elliptic coordinates
\begin{gather*}
\xi_{1,2}=\kappa+q_1^2+q_2^2\pm\sqrt{\kappa^2-2\big(q_1^2-q_2^2\big)\kappa+\big(q_1^2+q_2^2\big)^2}
\end{gather*}
up to the constant factor. In nonholonomic case, substituting the same tensor $K^{(4)}$ into the characteristic tensor $\hat{\mathbf K}= K^{(4)} \hat{\mathrm g}^{-1}$, one gets the following eigenvalues
\begin{gather*}
\hat{\xi}_{1,2}=\frac{\big(1+q_1^2\big)\kappa+q_1^2+q_2^2\pm
\sqrt{\big(1+q_1^2\big)^2\kappa^2-2\big(q_1^2\big(q_1^2+q_2^2\big)+q_1^2-q_2^2\big)\kappa+\big(q_1^2+q_2^2\big)^2}}{1+q_1^2+q_2^2}. \end{gather*}
Solutions of the compatibility equations~(\ref{meq-2h}) and~(\ref{meq-2}) are labelled by two arbitrary func\-tions~$F_1$ and~$F_2$ on these coordinates:
\begin{gather*}
V(q_1,q_2)=\frac{F_1(u_1)-F_2(u_2)}{u_1-u_2},\qquad u_{1,2}=\xi_{1,2}\qquad\mbox{or}\qquad u_{1,2}=\hat{\xi}_{1,2}.
\end{gather*}

\textit{Parabolic coordinates.}
In Hamiltonian case eigenvalues of the characteristic tensor $\mathbf K= K^{(3)} \mathrm g^{-1}$ are standard parabolic coordinates
\begin{gather}\label{par-coordh}
\zeta_{1,2}=q_1\pm\sqrt{q_1^2+q_2^2},
\end{gather}
whereas in nonholonomic case eigenvalues of the characteristic tensor $\hat{\mathbf K}= K^{(3)} \hat{\mathrm g}^{-1}$ read as
\begin{gather}\label{par-coord}
\hat{\zeta}_{1,2}=\frac{q_1\pm\sqrt{q_2^2+1}\sqrt{q_1^2+q_2^2}}{1+q_1^2+q_2^2}.
\end{gather}
It is easy to express new characteristic coordinates via standard ones
\begin{gather*}
\hat{\zeta}_{1,2}=\frac{2\big(\zeta_1+\zeta_2\pm(\zeta_1-\zeta_2)\sqrt{1-\zeta_1\zeta_2 }\big)}{(\zeta_1-\zeta_2)^2+4}.
\end{gather*}
In both cases the desired separable potentials are equal to
\begin{gather*}
V(q_1,q_2)=\frac{F_1(u_1)-F_2(u_2)}{u_1-u_2},\qquad u_{1,2}=\zeta_{1,2}\qquad\mbox{or}\qquad u_{1,2}=\hat{\zeta}_{1,2}.
\end{gather*}

\textit{Polar coordinates.} Using one nontrivial eigenvalue of characteristic tensors $\mathbf K= K^{(2)}\mathrm g^{-1}$ and $\hat{\mathbf K}= K^{(2)} \hat{\mathrm g}^{-1}$ we can introduce only one coordinate
\begin{gather*}
r=\sqrt{q_1^2+q_2^2},\qquad\mbox{or}\qquad \hat{r}=\sqrt{\frac{q_1^2+q_2^2}{1+q_1^2+q_2^2}}.
\end{gather*}
The second coordinate is a function on $q_1/q_2$, for instance, it could be an angle $\varphi=\arctan q_1/q_2$, because solutions of the compatibility equations~(\ref{meq-2h}) and~(\ref{meq-2}) have the following form
\begin{gather*}
V(q_1,q_2)=F_1(\rho)+\frac{F_2(\varphi)}{\rho^2},
\qquad \rho=r\qquad\mbox{or}\qquad \rho=\hat{r}.
\end{gather*}

 \textit{Cartesian coordinates.} Separable in the Cartesian coordinates solution of the compatibility condition~(\ref{meq-2h}) reads as
 \begin{gather}\label{sep-car}
 V(q_1,q_2)=F_1(q_1)+F_2(q_2).
 \end{gather}
In nonholonomic case characteristic tensor $\hat{\mathbf K}= K^{(1)} \hat{\mathrm g}^{-1}$ yields only one coordinate
\begin{gather*}
\hat{\varrho}=\sqrt{\frac{1+q_1^2}{1+q_1^2+q_2^2}}
\end{gather*}
and the separable solution of the compatibility condition (\ref{meq-2}) looks like
\begin{gather*}
V(q_1,q_2)=F_1(\hat{\varrho})+\frac{F_2(\hat{\phi})}{\hat{\varrho}^2},\qquad \hat{\phi}=\arctan q_1.
\end{gather*}
Thus, after nonholonomic reduction one gets separable potential, which is dif\/ferent, even in the form of standard potential~(\ref{sep-car}) separable in Cartesian coordinates.

Summing up, we have found four characteristic coordinate systems for the Bertrand--Darboux type equation (\ref{bd-non}) associated with the integrable non-Hamiltonian vector f\/ield~$\hat{X}$~(\ref{conf-X}). Additional f\/irst integral of this f\/ield $\hat{X}$ is a polynomial of the second order in momenta def\/ined by the standard Killing tensor on the plane. Consequently, we can directly apply the standard St\"{a}ckel theory to the nonholonomic Heisenberg type systems.

\section{St\"{a}ckel systems}\label{section3}
In Hamiltonian case there is a one-to-one correspondence between the so-called St\"{a}ckel systems, integrable Killing tensors which mutually commute in the algebraic sense and separation of variables \cite{ben93,eis34}.

The nondegenerate $n\times n$ St\"{a}ckel matrix $S$, which $j$ column
depends only on variable $u_j$
\begin{gather*}
\det S\neq 0,\qquad \frac{\partial S_{kj}}{\partial u_m}=0, \qquad j\neq m\end{gather*}
def\/ines $n$ functionally independent integrals of motion
\begin{gather}\label{st-int}
H_k=\sum_{j=1}^n C_{jk}\big(p_{u_j}^2+U_j(u_j)\big),\qquad C=S^{-1},
\end{gather}
which are in involution with respect to canonical Poisson brackets
\begin{gather*}
\{u_i,p_{u_j}\}=\delta_{ij},\qquad \{u_i,u_j\}=\{p_{u_i},p_{u_j}\}=0.
\end{gather*}
The common level surface of the f\/irst integrals $H_1=\alpha_1,\ldots, H_2=\alpha_n$ is dif\/feomorphic to the $n$-dimensional real torus and one immediately gets
\begin{gather*}
p_{u_j}^2= \sum_{k=1}^n \alpha_kS_{kj}(u_j)-U_j(u_j).
\end{gather*}
It allows us to calculate quadratures for the corresponding Hamiltonian vector f\/ield $X=PdH_1$:
\begin{gather}\label{jac-eq}
\sum_{j=1}^n\int_{\gamma_{0}(p_0,q_0)}^{\gamma_j(p_{u_j},u_j)}\frac{S_{kj}(u ) du }
{\sqrt{\sum\limits_{k=1}^n \alpha_k S_{kj}(u)-U_j(u )}}=\beta_k, \qquad k=1,\ldots,n,
\end{gather}
where $\beta_1=t$ and $\beta_2,\ldots,\beta_n$ are constants of integration. Solution of the problem is thus reduced to solving a sequence of one-dimensional problems, which is the essence of the method of separation of variables~\cite{ben93}.

\subsection{St\"{a}ckel matrices for reduced systems}
In the def\/inition of the St\"{a}ckel integrals of motion (\ref{st-int}) momenta $p_u$ have to be canonically conjugated to eigenvalues $u$ of the characteristic $(1,1)$ tensor~$\mathbf K$. For instance, if we take tensor~$K^{(3)}$~(\ref{4-ten}) the eigenvalues of $\mathbf K$ are parabolic coordinates~$\zeta_{1,2}$~(\ref{par-coordh}) and the corresponding momenta read as
\begin{gather*}
p_{{\zeta}_{1,2}}=\frac{p_1}{2}\pm\frac{\sqrt{q_1^2+q_2^2}\mp q_1}{2q_2} p_2.
\end{gather*}
In nonholonomic case the eigenvalues of $\hat{\mathbf K}$ are given by $\hat{\zeta}_{1,2}$ (\ref{par-coord}) and the conjugated momenta have the following form
\begin{gather*}
p_{\hat{\zeta}_{1,2}}=\frac{q_2^2+1\pm q_1\sqrt{1+q_2^2}\sqrt{q_1^2+q_2^2}}{2\big(1-q_1^2\big)}p_1-\frac{q_1\mp\sqrt{1+q_2^2}\sqrt{q_1^2+q_2^2}}{2q_2}p_2.
\end{gather*}
In similar manner we can calculate all the canonical variables associated with tensors~$K^{(j)}$ (\ref{4-ten}) and to obtain standard St\"{a}ckel matrices in Hamiltonian case
\begin{gather*}
S^{(1)}=\left(
 \begin{matrix}
 1 & 1 \\
 1 & -1 \\
 \end{matrix}
 \right),\qquad
S^{(2)}=\left(
 \begin{matrix}
 0 & 1 \\
 1 & -r^{-2} \\
 \end{matrix}
 \right),\qquad
S^{(3)}=\left(
 \begin{matrix}
 1 & 1 \\
 -\zeta_1^{-1} & -\zeta_2^{-1}
 \end{matrix}
 \right)
\end{gather*}
and
\begin{gather*}
S^{(4)}=\left(
 \begin{matrix}
 \dfrac{1}{\xi_1-\kappa^2} & \dfrac{1}{\xi_2-\kappa^2} \vspace{1mm}\\
 \dfrac{-1}{\xi_1(\xi_1-\kappa^2)} & \dfrac{-1}{\xi_2(\xi_2-\kappa^2)}
 \end{matrix} \right).
\end{gather*}
For the nonholonomic Heisenberg type systems associated with tensors $K^{(j)}$ (\ref{4-ten}) St\"{a}ckel mat\-rices are equal to
\begin{gather*}
\hat{S}^{(1)}=\left(
 \begin{matrix}
 0 & \dfrac{1}{\hat{\varrho}^2(1-\hat{\varrho}^2)}\vspace{1mm}\\
 1 & \dfrac{-1}{\hat{\varrho}^4(1-\hat{\varrho}^2)}
 \end{matrix}
 \right),\qquad
\hat{S}^{(2)}=\left(
 \begin{matrix}
 0 & \dfrac{1}{1-\hat{r}^2} \vspace{1mm}\\
 1 & \dfrac{-1}{\hat{r}^2(1-\hat{r}^2)}
 \end{matrix}
 \right)
\end{gather*}
and
\begin{gather*}
\hat{S}^{(3)}=
\left(
 \begin{matrix}
 \dfrac{1}{1-\hat{\zeta}_1} & \dfrac{1}{1-\hat{\zeta}_2} \vspace{1mm}\\
 \dfrac{-1}{\hat{\zeta}_1(1-\hat{\zeta}_1)} & \dfrac{-1}{\hat{\zeta}_2(1-\hat{\zeta}_2)}
 \end{matrix}
 \right),\\ \hat{S}^{(4)}=\left(
 \begin{matrix}
 \dfrac{1}{(\hat{\xi}_1-\kappa^2)(1-\hat{\xi}_1)} & \dfrac{1}{(\hat{\xi}_2-\kappa^2)(1-\hat{\xi}_2)} \vspace{1mm}\\
 \dfrac{-1}{\hat{\xi}_1(\hat{\xi}_1-\kappa^2)(1-\hat{\xi}_1)} & \dfrac{-1}{\hat{\xi}_2(\hat{\xi}_2-\kappa^2)(1-\hat{\xi}_2)}
 \end{matrix}
 \right).
\end{gather*}
It is easy to see, that imposing linear, non integrable constraint~(\ref{f-nhol}) in polar, parabolic and elliptic cases we have to multiply $j$ column of the standard St\"{a}ckel matrices on the function~$(1-u_j)$, where~$u_j$ is the eigenvalue of $\hat{\mathbf K}$. Similar transformations of the St\"{a}ckel matrices for nonholomic systems on a two-dimensional sphere are discussed in~\cite{ts12}.

Using a new time variable $\tau$ def\/ined by
\begin{gather*}
dt=\mu^{-1}d\tau\equiv\big(1+q_1^2+q_2^2\big)d\tau
\end{gather*}
we can rewrite the conformally Hamiltonian vector f\/ield $\hat{X}$ (\ref{conf-X}) in the Hamiltonian form and obtain quadratures similar to~(\ref{jac-eq}).

Thus, in nonholonomic case solution of the problem is reduced to solving a sequence of one-dimensional problems after suitable change of time.

\subsection{Compatible Poisson brackets}
Riemannian geometry is not, a priori, concerned with symplectic or Poisson structures. Nevertheless, it is known that any tensor f\/ield $L$ with vanishing Nijenhuis torsion on the Riemannian manifold~$Q$ yields trivial deformation of the canonical Poisson bracket $\{\cdot,\cdot\}$ on its cotangent bundle $T^*Q$
\begin{gather}\label{tur-def}
\{q_i,q_j\}_L=0,\qquad \{q_i,p_j\}_L=-L^{ij},\qquad \{p_i,p_j\}_L=\left(\frac{\partial L^{kj}}{\partial q_i}-\frac{\partial L^{ki}}{\partial q_j}\right)p_k,
\end{gather}
where $(p_i,q_i)$ are f\/ibered coordinates. This Poisson bracket is compatible with the canonical one, so that there is a recursion operator $N=P_LP^{-1}$, which allows us to construct a whole family of compatible brackets associated with Poisson bivectors $P_L^{(m)}=N^mP$ on $T^*Q$ \cite{ts11,ts12}.

According to \cite{ben93} generic solution of the Killing equation $K$ (\ref{gen-K}) determines a conformal Killing tensor with vanishing Nijenhuis torsion
\begin{gather*}
L=K-\operatorname{tr}(K) \mathrm g
\end{gather*}
and Turiel's deformation (\ref{tur-def}) of the canonical Poisson bracket \cite{ts11,ts12}. The eigenvalues of the corresponding recursion operator $N=P_LP^{-1}$ are characteristic coordinates. Integrals of mo\-tion~$H_1$~(\ref{h1-hol}) and~$H_2$~(\ref{h2-hol}) are in involution with respect to the compatible Poisson brackets
\begin{gather*}
\{H_1,H_2\}=\{H_1,H_2\}_L=0
\end{gather*}
and to other polynomial Poisson brackets associated with the Poisson bivectors~$P_L^{(m)}$. Indeed, we can f\/ind such $L$-tensors directly from the Hamilton function~$H_1$ using modern software~\cite{ts05}.

For the Heisenberg type systems, we have to solve the same Killing equation~(\ref{meq-1}) and second equation on potential with~$(1,1)$ tensor $\hat{\mathbf K}=K\hat{\mathrm g}$. This tensor does not appear in the Turiel construction of the Poisson bracket and we can directly prove the following proposition.
\begin{proposition}
Integrals of motion $\hat{H}_1$ \eqref{h1-nhol} and $\hat{H}_2$ \eqref{h2-nhol} are in involution
\begin{gather*}
\big\{\hat{H}_1,\hat{H}_2\big\}^\wedge=\big\{\hat{H}_1,\hat{H}_2\big\}_L=0
\end{gather*}
with respect to the compatible Poisson brackets $\{\cdot,\cdot \}^\wedge$~\eqref{nhcan-br} and $\{\cdot,\cdot \}_L$~\eqref{tur-def}. The eigenvalues of the recursion operator \begin{gather*}\hat{N}=P_L\hat{P}^{-1},\end{gather*}
 where $\hat{P}$ is given by~\eqref{conf-X}, are characteristic coordinates discussed in the previous Section.
\end{proposition}
The proof consists of the straightforward calculations.

Thus, starting with a common Poisson bivector $P_L$ we can construct two recursion opera\-tors~$N$ and~$\hat{N}$ and two families of the Poisson brackets on~$T^*\mathbb R^2$.

\subsection{Lax matrices}
We can construct Lax representations for the St\"{a}ckel systems with uniform rational and polynomial potentials $U_j=U$ in~(\ref{st-int}), see \cite{ts94} and references within. For the sake of brevity we consider only systems associated with tensor~$K^{(3)}$~(\ref{4-ten}).

If $V(q_1,q_2)=0$, we can introduce a well-known $2\times 2$ Lax matrix for the geodesic motion on the plane
\begin{gather}\label{lax1}
\mathcal{L}(u)=\left(
 \begin{matrix}
 h& e\\
 f & -h
 \end{matrix}
 \right)(u)\equiv\left(
 \begin{matrix}
 \dfrac{1}{2}\dfrac{d e(u)}{dt} & e(u) \vspace{1mm}\\
 - \dfrac{1}{2}\dfrac{d^2 e(u)}{dt^2} & - \dfrac{1}{2}\dfrac{d e(u)}{dt}
 \end{matrix}
 \right),
\end{gather}
which satisf\/ies to the Lax equation
\begin{gather}\label{eq-lax}
\frac{d}{dt} \mathcal L=[\mathcal L,\mathcal A]\equiv\mathcal L\mathcal A-\mathcal A\mathcal L,
\qquad\text{with}\qquad \mathcal A=\left(
 \begin{matrix}
 0 & 1 \\
 0 & 0
 \end{matrix}
 \right).
\end{gather}
This equation guarantees that the eigenvalues of $\mathcal L$ are conserved quantities in the involution.

In our case function $e(u)$ depends on the so-called spectral parameter~$u$
\begin{gather*}
e(u)=\frac{(u-\zeta_1)(u-\zeta_2)}{u}=u-2 q_1-\frac{q_2^2}{u},
\end{gather*}
which is completely def\/ined by the eigenvalues of the characteristic tensor $\mathbf K= K^{(3)}\mathrm g^{-1}$ or parabolic coordinates~$\zeta_{1,2}$ on the plane.

The involution property of eigenvalues of the Lax matrix $\mathcal L$ is equivalent to existence of a~classical $r$-matrix~$r_{12}$, that as
\begin{gather*}
\{\mathcal L_1(u), \mathcal L_2(v)\}=[r_{12}(u,v),\mathcal L_1(u)]-[r_{21}(u,v),\mathcal L_2(v)] .
\end{gather*}
Here we use the familiar notation for tensor product of $\mathcal L$ and unit matrix $I$
\begin{gather*}\mathcal L_{1}(u)=\mathcal L(u)\otimes \mathrm I,\qquad \mathcal L_2(v)=\mathrm I\otimes\mathcal L(v),\qquad r_{21}(u,v)=\Pi r_{12}(v,u) \Pi,\end{gather*}
whereas $\Pi$ is the permutation operator: ${\Pi}x\otimes y =y\otimes x$, $\forall\, x,y$. Evaluating canonical Poisson brackets~(\ref{can-br}) between entries of the Lax matrix~$\mathcal L(u)$~(\ref{lax1}) we obtain a~constant $r$-matrix
\begin{gather}\label{r-1}
r_{12}(u,v)=\frac{2}{v-u} \Pi
\equiv\frac{2}{v-u}
\left(\begin{matrix}
 1 & 0 & 0 & 0 \\
 0 & 0 & 1 & 0 \\
 0 & 1 & 0 & 0 \\
 0 & 0 & 0 & 1
\end{matrix}\right).
\end{gather}

If we want to consider separable in parabolic coordinates potentials $V(q_1,q_2)$, we have to add some items to the initial Lax matrix $\mathcal L(u)$~(\ref{lax1}). For instance, we can take the following additive perturbation
\begin{gather*}
\mathcal L_V(u)=\mathcal L(u)+\left(
 \begin{matrix}
 0 & 0 \\
 \Delta f & 0
 \end{matrix}
 \right),\qquad \Delta f=\big[\phi(u)e^{-1}(u)\big],
\end{gather*}
where $\phi(u)$ is an arbitrary constant function and $ [g(u) ]$ is a truncated Laurent series expansion of~$g(u)$ with respect to the variable $u$. For instance, when $\phi(u)=-au^3$ the Taylor part of expansion about $u=\infty$ looks like
\begin{gather*}
\big[{-}au^3 e^{-1}(u)\big]=-a\big(u^2+2uq_1+q_2^2+4q_1^2\big).
\end{gather*}
Substituting this expression into $\mathcal L_V(u) $ one gets a Lax matrix for the H\'{e}non--Heiles system on the plane with the Hamiltonian
\begin{gather*}
H_1=\frac{1}{2}\big(p_1^2+p_2^2\big)+8aq_1\big(2q_1^2+q_2^2\big).
\end{gather*}

In nonholonomic case we can also f\/ind similar Lax matrices, classical $r$-matrix and integrable potentials.
\begin{proposition}
For the Heisenberg type systems equations of motion \eqref{conf-X} when $V(q_1,q_2)=0$ can be rewritten in the Lax form~\eqref{eq-lax} if
\begin{gather}\label{lax2}
\hat{\mathcal{L}}(u)=\left(
 \begin{matrix}
 \hat{ h}& \hat{e}\\
 \hat{f} & -\hat{h}
 \end{matrix}
 \right)(u)\equiv\left(
 \begin{matrix}
 \dfrac{1}{2\mu}\dfrac{d\hat{e}(u)}{dt} & \hat{e}(u) \vspace{1mm}\\
 - \dfrac{d}{\mu dt}\left(\dfrac{1}{2\mu}\dfrac{d \hat{e}(u)}{dt}\right) -2\hat{H}_1 \hat{e}(u)& - \dfrac{1}{2\mu}\frac{d \hat{e}(u)}{dt}
 \end{matrix}
 \right)
 \end{gather}
and
\begin{gather*}
 \hat{\mathcal A}=\left(
 \begin{matrix}
 0 & \mu \\
 -2\mu \hat{H}_1 & 0
 \end{matrix}
 \right).
\end{gather*}
Here $\mu=\big(1+q_1^2+q_2^2\big)^{-1}$ is a conformal factor~\eqref{conf-X} and function
\begin{gather*}
\hat{e}(u)=\frac{\big(u-\hat{\zeta}_1\big)\big(u-\hat{\zeta}_2\big)}{u}=u-\frac{2q_1}{1+q_1^2+q_2^2}
-\frac{q_2^2}{u\big(1+q_1^2+q_2^2\big)}
\end{gather*}
is completely defined by characteristic coordinates $\hat{\zeta}_{1,2}$~\eqref{par-coord}.
\end{proposition}

The Lax matrix (\ref{lax2}) was constructed by using the generic construction of the Lax matrices for the St\"{a}ckel systems~\cite{ts94}. The modif\/ication consists only of application of the nontrivial conformal factor~$\mu$.

Evaluating Poisson brackets $\{\cdot,\cdot\}^\wedge$ (\ref{nhcan-br}) between entries of the Lax matrix $\hat{\mathcal L}(u)$ (\ref{lax2}) we f\/ind the corresponding classical $r$-matrix
\begin{gather}\label{r-2}
\hat{r}_{12}(u,v)=r_{12}(u,v)+2\left(\begin{matrix}
 0 & 0 & 0 & 0 \\
 0 & 0 & \hat{e}(v) & 0 \\
 0 & 0 & 0 & 0 \\
 -\hat{f}(v) & 0 & 0 & 0
\end{matrix}\right),
\end{gather}
which explicitly depends on dynamical variables via entries of $\hat{\mathcal L}(u)$ (\ref{lax2}).

As above, we can consider additive perturbation of this Lax matrix
\begin{gather*}
\hat{\mathcal L}_V(u)=\hat{\mathcal L}(u)+\left(
 \begin{matrix}
 0 & 0 \\
 \Delta\hat{f} & 0
 \end{matrix}
 \right),\qquad \Delta \hat{f}=\big(u^2-1\big)\big[\phi(u)\hat{e}^{-1}(u)\big]
\end{gather*}
in order to get Lax matrices for the nonholonomic Heisenberg type systems with~$V\neq 0$. For instance, when $\phi(u)=-au^3$ the Teylor part of expansion about $u=\infty$ looks like
\begin{gather*}
\big[{-}au^3 \hat{e}^{-1}(u)\big]=-a\left(u^2+\frac{2uq_1}{1+q_1^2+q_2^2}+\frac{q_2^2}{{1+q_1^2+q_2^2}}+\frac{4q_1^2}{\big(1+q_1^2+q_2^2\big)^2}\right).
\end{gather*}
Substituting this expression into $\hat{\mathcal L}_V(u)$ one gets a Lax matrix for the nonholonomic counterpart of the H\'{e}non--Hieles system on the plane with the Jacobi integral
\begin{gather*}
\hat{H}_1=\frac{1}{2}\big(p_1^2+p_2^2+(q_2p_1-q_1p_2)^2\big)+\frac{8aq_1\big(q_1^2q_2^2+q_2^4+2q_1^2+q_2^2\big)}{\big(1+q_1^2+q_2^2\big)^3}.
\end{gather*}

\section{Conclusion}\label{section4}
Imposing nonholonomic constraints to Hamiltonian systems and making a suitable reduction one gets some special class of non Hamiltonian systems on the reduced phase space, see book~\cite{bl03}. Usually we can not investigate the reduced system using standard mathematical methods of Hamiltonian dynamics. For instance, we do not know how to get Lax matrices, classical $r$-matrices or compatible Poisson brackets for the conformally Hamiltonian systems associated with the nonholonomic Chaplygin ball, nonholonomic Suslov or Veselova systems.

In this note we f\/ind Killing tensors and compatible Poisson brackets, describe integrable potentials and characteristic coordinates, evaluate St\"{a}ckel matrices and St\"{a}ckel quadratures, to show Lax matrices and classical $r$-matrices for the nonholonomic Heisenberg type systems. Indeed, we prove that some modern geometric methods of Hamiltonian mechanics can be directly applied to the reduced conformally Hamiltonian systems. Other similar examples can be found in \cite{bts12,bm15,ts11a,ts12,ts15b}.

Of course, we can impose other nonholonomic constraints on the original Hamiltonian system~(\ref{eq-3dh}),~(\ref{h-free}). It may be interesting to describe all the constraints which lead to nontrivial integrable metrics and potentials on the reduced phase space.

\subsection*{Acknowledgements}
We are very grateful to the referees for thorough analysis of the manuscript, constructive suggestions and proposed corrections, which certainly lead to a~more profound discussion of the results. We are also deeply grateful A.V.~Borisov and I.A.~Bizayev for the relevant discussion.

Section~\ref{section2} was written by A.V.~Tsiganov and supported by the Russian Science Foundation (project~15-12-20035).
Section~\ref{section3} was written by Yu.A.~Grigoryev and A.P.~Sozonov within the framework of the Russian Science Foundation (project
 15-11-30007).

\pdfbookmark[1]{References}{ref}
\LastPageEnding

\end{document}